\documentstyle[psfig,floats,aps,twocolumn]{revtex}

\def\al{CBr$_{4}$-C$_{2}$Cl$_{6}$}
\def\mis{~$\mu ms^{-1}$}

\begin{document}

\draft\twocolumn[\hsize\textwidth\columnwidth\hsize\csname
@twocolumnfalse\endcsname

\title{Pattern Stability and Trijunction Motion in Eutectic Solidification}

\author{S. Akamatsu$^1$, M. Plapp$^2$, G. Faivre$^1$, and A. Karma$^3$}

\address{$^1$Groupe de Physique des Solides, CNRS UMR 7588,
         Universit\'es Denis-Diderot et Pierre-et-Marie-Curie, Tour 23,
         2 place Jussieu, 75251 Paris Cedex 05, France\\
              $^2$Laboratoire de Physique de la Mati\`ere Condens\'ee,
                  CNRS/Ecole Polytechnique, 91128 Palaiseau, France\\
              $^3$Physics Department, Northeastern University,
                  Boston, Massachusetts 02115}

\date{February 28, 2002}

\maketitle

\begin{abstract}
We demonstrate by both experiments and phase-field simulations that
lamellar eutectic growth can be stable for a wide range of spacings
below the point of minimum undercooling at low velocity, contrary to
what is predicted by existing stability analyses.  This
overstabilization can be explained by relaxing Cahn's assumption that
lamellae grow locally normal to the eutectic interface.
\end{abstract}

]

\begin{figure}
\centerline{ \psfig{file=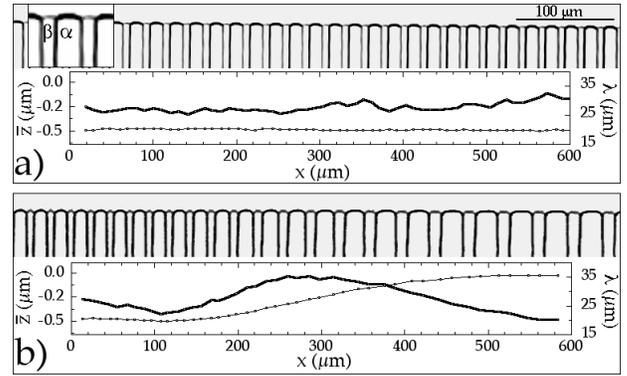,width=.45\textwidth}} \medskip
\caption{Photographs: Lamellar-eutectic fronts of a nearly eutectic
CBr$_4$-C$_2$Cl$_6$ alloy in directional solidification (the growth
direction is upward) in $12$-$\mu m$ thick samples.  Graphs:
Interlamellar spacing $\lambda$ (thin lines) and position $\bar z$ of
the front (thick lines) as functions of the space variable $x$.  a)
Stationary pattern ($V = 0.5$\mis; $G= 80 Kcm^{-1}$).  b) Modulated
pattern ($V = 0.25$\mis; $G= 48 Kcm^{-1}$).}
\label{arrays}
\end{figure}

The solidification of eutectic alloys is both a striking example of
spontaneous pattern formation in nature and a metallurgical problem of
widely recognized practical importance \cite{KurFis}.  This growth
process has been traditionally studied by directional solidification
experiments where a sample containing a binary alloy of near-eutectic
composition is pulled with a fixed speed $V$ in an externally imposed
temperature gradient $G$.  This setup produces a wide range of
microstructures of which the simplest is an array of lamellae of two
coexisting $\alpha$ and $\beta$ solid phases growing into the
metastable liquid, as shown in Figure \ref{arrays}a.  The steady-state
growth of a perfectly periodic lamellar array is described by the
classic Jackson-Hunt (JH) theory \cite{Jackson66} that predicts the
relationship
\begin{equation}
\Delta T(\lambda)\equiv T_E- T_{av}(\lambda)=K_1V\lambda+K_2/\lambda,
\label{JH}
\end{equation}
between the lamellar spacing $\lambda$ (width of one lamella pair) and
the undercooling, $\Delta T(\lambda)$, which is the difference between
the eutectic temperature, $T_E$, at which the three phases ($\alpha$,
$\beta$ and liquid) coexist in equilibrium, and the average
temperature of the nonequilibrium eutectic interface spatially
averaged over one spacing, $T_{av}(\lambda)$.  The first and second
terms on the right-hand-side of Eq.  \ref{JH} represent the
undercooling necessary to drive the diffusive transport of the two
chemical components of the alloy in the liquid (coupled growth) and
the capillary undercooling associated with the curvature of the
solid-liquid interface, respectively; $K_1$ and $K_2$ are constants
that only depend on the alloy system and the overall composition of
the sample.

The JH result implies that the growth
undercooling has a minimum value
$\Delta T_m=(4K_1K_2)^{1/2}\,V^{1/2}$
for a spacing
\begin{equation}
\lambda_m=(K_2/K_1)^{1/2}\,V^{-1/2},\label{sel1}
\end{equation}
which is easily found by setting $d\Delta T/d\lambda=0$.
%---------------------figure-----------------
%--------------------------------------------

Lamellar growth is well known to be unstable for spacings smaller than
a critical value $\lambda_c$.  This instability leads to the local
elimination of lamellae and is the mechanism by which the array
increases its average spacing during the dynamical transient that
produces the final pattern.  Hence, it is crucially important for
understanding pattern selection in this system.  Oscillatory
instabilities are also known to limit the array stability at large
spacing \cite{Karma96,Ginibre97}.  JH have credited Cahn in Ref.
\cite{Jackson66} for pointing out that $\lambda_c$ should be equal to
$\lambda_m$ if one assumes that lamellae grow locally normal to the
envelope of the eutectic interface.  Langer later formalized this
result by showing that a large-scale and small-amplitude spacing
modulation of a steady-state array obeys the diffusion equation
\cite{Langer80}
\begin{equation}
\partial_t \lambda(x,t) =D\,
\partial_x^2 \lambda(x,t),\label{diff}
\end{equation}
where $x$ is the coordinate
perpendicular to the growth axis $z$,
and $D=D_\perp$ with
\begin{equation}
D_\perp=\frac{V\lambda_0}{G}\left.
\frac{d\Delta T(\lambda)}{d\lambda}\right|_{\lambda=\lambda_0}
=\frac{K_{1}\lambda_0V^2}{G}
\left(1-\frac{1}{\Lambda_0^2}\right).\label{Dperp}
\end{equation}
We have defined $\Lambda_0 = \lambda_0/\lambda_m$ where $\lambda_0$ is
the spacing of the steady-state array being perturbed, and we have
used the subscript ``$\perp$'' to stress that this expression for $D$
is obtained under Cahn's assumption that lamellae grow normal to the
interface. Langer's analysis reproduces the JH-Cahn result that
growth is unstable below $\lambda_{m}$ since perturbations are
amplified (decay) when $D_\perp<0$ ($>0$).  In addition, it shows that
lamella elimination is initiated by a long-wavelength diffusive
instability that is generically present in one-dimensional pattern
forming systems with translation symmetry along $x$.

In this letter, we study the steady-state and stability
properties of lamellar eutectic growth by thin-sample
directional solidification experiments in the
transparent organic system CBr$_4$-C$_2$Cl$_6$ and
by two-dimensional simulations of
a phase-field model, and we extract from both approaches
independent accurate determinations of $\lambda_m$
and $\lambda_c$. An important and novel component of our
experiments is the direct
measurement of $\Delta T(\lambda)$,
which allows us to obtain $\lambda_m$ from the minimum of
this curve rather than computing its value from
Eq. \ref{sel1}, thereby circumventing uncertainties in
materials parameters.
We find that, in both experiments and
simulations, $\lambda_c$ is substantially smaller than $\lambda_m$,
even for typical directional solidification growth
conditions where the two spacings have previously
been assumed equal. Furthermore, by analyzing the
decay of long-wavelength perturbations of the array in
both experiments and simulations, we obtain a direct measurement
of $D$, which allows us to shed light
on the origin of the discrepancy between $\lambda_c$
and $\lambda_m$.

The experiments were made with a nearly eutectic \al~alloy prepared with
zone refined materials in thin ($12~\mu m$ thick) glass wall samples
($8~mm$ wide and $60~mm$ long). The values of $G$ used ranged from
$40~Kcm^{-1}$ to $110~Kcm^{-1}$ ($\pm 10 \%$), and those of $V$ from
0.125 to 0.75\mis ($\pm 4 \%$). Details concerning the preparation of
the samples, the solidification setup, and the visualization of the
front shape can be found in Refs.~\cite{Ginibre97,AkaFai00}.

\begin{figure}
\centerline{ \psfig{file=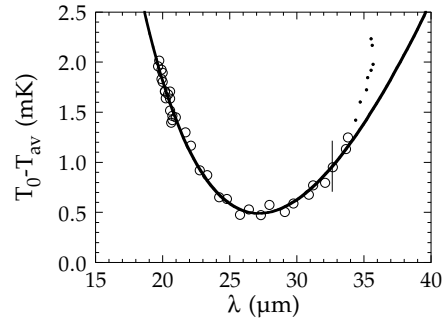,width=.33\textwidth}} \medskip
\caption{Undercooling $T_{0}-T_{av}$ vs spacing $\lambda$ measured
experimentally from the photograph shown in Fig.~1b (open circles and
dots). Thick line: Best fit of the JH law, Eq.~\protect\ref{JH},
to the data represented by open circles.
Vertical bar: error range. A remarkable feature 
is the presence of spacings considerably smaller than 
$\lambda_m\approx 27\mu m$. The value of the smallest
stable spacing $\lambda_c$ predicted from 
Eq.~\protect\ref{pred} below is $19~\mu m$.}
\label{under}
\end{figure}

The steady-state $\Delta T(\lambda)$ curve has never been measured
directly due to the fact that $\Delta T_m$ is usually of the order
$0.01\,K$, whereas the absolute temperature is not known with a
precision better than about $0.1~K$.  To overcome this difficulty, we
exploit two key ingredients.  Firstly, as will be described elsewhere,
we are able to create a large-scale modulation of spacing where
$\lambda(x)$ varies between two extreme values that comprise
$\lambda_m$, as shown in Fig.  \ref{arrays}(b).  Secondly, we measure
the $z$ coordinate of the solid-liquid interface averaged over one
$\lambda$, which we denote by $\bar z(x)$, and compute the local front
undercooling using $T_{av}(x)=G\bar z(x)+T_0$, where $T_0$ is an
unknown constant.  By eliminating $x$ between $T_{0}-T_{av}(x)$ and
$\lambda(x)$, we obtain $T_{0}-T_{av}(\lambda)$ which we then fit to
Eq.~\ref{JH} expressed in the form $T_{0}-T_{av}(\lambda)=\Delta
T_m(\lambda/\lambda_m+\lambda_m/\lambda)/2 - \Delta T_{0}$, using
$\lambda_m$, $\Delta T_m$ and $\Delta T_{0} = T_{E}-T_{0}$ as
adjustable parameters.  A plot of $T_{0}-T_{av}(\lambda)$ and its fit
is shown in Fig.~\ref{under}.  The fit is very good for $\lambda$
smaller than about $1.25\lambda_{m}$.  The departure observed beyond
this limit is compatible with the one which exists between the
numerically calculated curves $\Delta T(\lambda)$ and the JH
approximation \cite{Karma96,kassmis}.  We performed such measurements
for $V$ ranging from 0.125 to 0.5 \mis.  We found
$\lambda_m^{2}V=K_2/K_1 = 193 \pm 16~\mu m^{3}s^{-1}$ and $\Delta
T_m^2/V = 4K_{1}K_{2}=(2.7 \pm 1.3) \times 10^{-3} K^{2}s\mu m^{-1}$.
These values compare well to those calculated from the material
constants of \al~given in ref.  \cite{MerFai93}, namely,
$\lambda_m^{2}V= 185 \pm 20~\mu m^{3}s^{-1}$ and $\Delta T_m^2/V =
(1.2 \pm 0.2) \times 10^{-3} K^{2}s\mu m^{-1}$.

%---------------------figure-----------------
%--------------------------------------------

To study experimentally the small spacing stability limit, we exploit
the fact that $\lambda_m\sim V^{-1/2}$. Therefore,
we can effectively vary $\lambda_0/\lambda_m$
by performing downward velocity jumps of relatively large
amplitude. Namely, we start from a stable quasi-stationary
periodic array of spacing $\lambda_0$ at a higher velocity,
and then observe whether the same array at
the lower velocity, and hence smaller $\lambda_0/\lambda_m$,
remains stable or becomes unstable.

\begin{figure}
\centerline{ \psfig{file=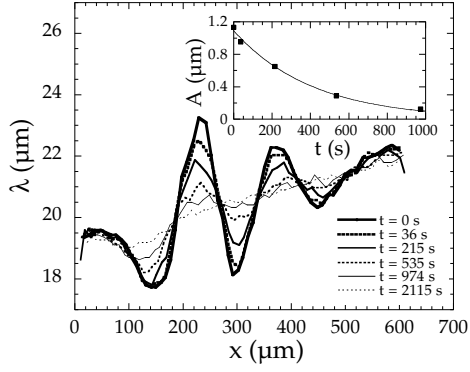,width=.35\textwidth}} \medskip
\caption{Experimental measurements of the spacing $\lambda$ vs the
space variable $x$ at different times $t$ showing the relaxation of a
large-scale modulation of a lamellar pattern of $\Lambda_{0} \approx
1$.  ($G= 80 Kcm^{-1}$;$V = 0.5$\mis).  Inset: Amplitude $A$ of the
dominant mode (wavelength $145 \mu m$) as a function of $t$, fitted by
an exponential law (time constant $410 s$).}
\label{relax}
\end{figure}

To measure experimentally the array diffusion constant
$D$, we use the fact that the amplitude of a
long-wavelength modulation
of spacing of the form
\begin{equation}
\lambda(x,t)\approx \lambda_0
+\delta \lambda_0 \exp(ikx+\omega_kt)\label{pertanal}
\end{equation}
decays exponentially in time when the array
is stable ($\lambda_0 > \lambda_c$).
Substituting this form in Eq. \ref{diff}, we obtain
the simple dispersion relation
$\omega_k=-Dk^2$
which is valid if the wavelength
of the perturbation $2\pi/k \gg \lambda_0$. Knowing
$k$, and calculating $\omega_k$ by a fit of the
measured amplitude of the modulation to a decaying
exponential,
we obtain $D$. This is illustrated in Fig. \ref{relax}
for a case where $2\pi/k \approx 7\lambda_0$.
We deduce from this measurement that
\begin{equation}
D=D_\perp+D_\parallel\label{Dnew}
\end{equation}
where $D_\parallel$ is a positive contribution responsible for the
overstabilization of the array.  The latter is calculated by taking
the difference between $D$, and $D_\perp$ evaluated via Eq.
\ref{Dperp} using the value $K_{1}= 1.9 \times 10^{-3} Ks\mu m^{-2}$
obtained from our present experiments.
%---------------------figure-----------------
%--------------------------------------------

Next, we simulate a phase-field model of a two-component (AB) eutectic
alloy which is completely symmetric under the exchange of $\alpha$ and
$\beta$. Our goal here is not to model quantitatively
the experiments but to demonstrate the generality of our results in
different alloy systems.  An order parameter $\phi$, which
distinguishes between solid ($\phi=+1$) and liquid ($\phi=-1$), is
coupled to the dimensionless concentration field, $u\equiv
(c-c_E)/(\Delta c/2)$, where $c$ is the mole fraction of B, and $\Delta
c$ is the difference between the solid composition in $\beta$ and
$\alpha$ at $T_E$.  The non-conserved and conserved dynamics for
$\phi$ and $u$, $\tau\partial_t\phi=-\delta F/\delta \phi$, and
$\partial_tu=\vec\nabla\cdot(M(\phi)\vec\nabla \delta F/\delta u)$,
respectively, are derived from the functional
\begin{equation}
F=\int_V dV\left[\frac{W_u^2}{2}|\nabla u|^2+
\frac{W_\phi^2}{2}|\nabla \phi|^2
+f(\phi,u,T)\right]
\end{equation}
where $V$ is the volume of the system and
\begin{equation}
f(\phi,u,T)=f_{DW}(\phi)+\frac{1+h(\phi)}{2}f_s
+\frac{1-h(\phi)}{2}f_l,
\end{equation}
is the bulk free-energy density that
is the sum of the standard double-well
$f_{DW}=-\phi^2/2+\phi^4/4$,
and a concentration-dependent part that interpolates between
the bulk free-energy of the liquid, $f_l=u^2/2$, and the solid
$f_s=(u^2-1)^2/8-(T_E-T)/T_E$,
with $h(\phi)=3(\phi-\phi^3/3)/2$. Furthermore, we used
the mobility $M(\phi)=D_l[1-(1+\phi)^4/16]$,
where $D_l$ is the solute diffusivity in the liquid, that
makes the diffusivity in the solid vanish and yields
efficient computations \cite{Plapp01}. Directional
growth is implemented using the frozen-temperature
approximation $T(z,t)=T_E+Gz-Vt$, and periodic boundary
conditions in $x$ are used in all simulations.

%---------------------figure-----------------
\begin{figure}
\centerline{ \psfig{file=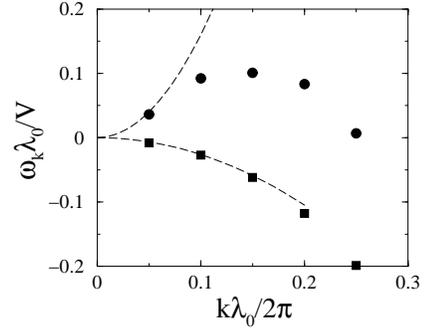,width=.3\textwidth}} \medskip
\caption{Growth rates $\omega_k$ of spacing perturbations versus wave
vector $k$ as extracted from phase-field simulations.  We took
$W_u=W_\phi=D_l=\tau=T_E=1$ and $V=0.01$, which gave
$\lambda_m/l_D=0.333$, where $l_D=D_l/V$ is the diffusion length.  We
measure the strength of the temperature gradient by the ratio
$l_T/l_D$, where $l_T=m\Delta c/G$ is the thermal length ($m\Delta c
=2$ in the phase-field units).  For both arrays, $\Lambda_0=0.84$;
circles: $l_T/l_D=20$; squares: $l_T/l_D=2$.  Dashed lines: fits
$\omega_k=Dk^2$ obtained from the points with smallest $k$.}
\label{stab}
\end{figure}
\begin{figure}
\centerline{ \psfig{file=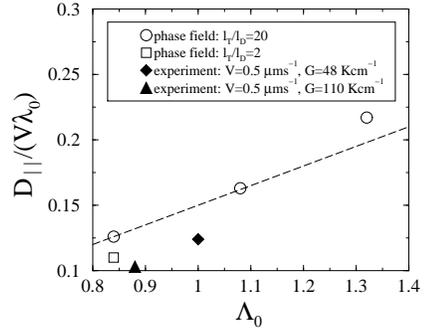,width=.3\textwidth}} \medskip
\caption{$D_\parallel/(V\lambda_0)$ versus $\Lambda_0$ for simulations
and experiments.  Dashed line: Eq.~\protect\ref{Dpar} drawn with
$A=0.15$.}
\label{diffcte}
\end{figure}

%--------------------------------------------
Steady-state $\Delta T(\lambda)$ curves were obtained from short
simulations with two lamellae.
The stability was studied with long simulations where steady-state
arrays of up to 20 lamellae, constructed from the two-lamella
solutions, are slightly perturbed by a long-wavelength modulation of
spacing of the form of Eq.~\ref{pertanal}.  By exponential fits of
the amplitude of modulation vs time for different $k$, we obtain the
stability spectrum $\omega_k$ as illustrated in Fig.~\ref{stab}, and
hence $D$ by a quadratic fit of $\omega$ vs $k$ at small $k$.  The
stability limit $\lambda_c$ is then obtained by determining where $D$
changes sign and values of $D_\parallel$ are obtained by subtracting
$D_\perp$ from $D$. We find that the dimensionless ratio
$D_\parallel/(V\lambda_0)$ varies with $\Lambda_0$, but negligibly
with $G$ and $V$.  Moreover, we find that the simple form
\begin{equation}
D_\parallel/(V\lambda_0)\approx  A \, \Lambda_0,\label{Dpar}
\end{equation}
with $A\approx 0.15$ gives a reasonable fit to
our phase-field simulation results
as shown in Fig. \ref{diffcte}. Remarkably, our experimentally
determined values of $D_\parallel$ are close to those obtained
in the phase-field simulations, even though the two alloy systems
are different.
%---------------------figure-----------------
%--------------------------------------------

To interpret our findings, let us briefly review
Langer's analysis that yields $D=D_{\perp}$. Its
first ingredient is the assumption
that the interface adjusts adiabatically
its average temperature to the local spacing, or
\begin{equation}
\Delta T[\lambda(x,t)]\approx -G\zeta(x,t),\label{adia}
\end{equation}
where $\Delta T(\lambda)$ is the same as in steady-state and
$\zeta(x)$ is the $z$ coordinate of the envelope of the eutectic
interface, defined as a smooth curve interpolating the positions
of the three-phase junctions (trijunctions), with the origin at $T_E$.
The second is Cahn's assumption that lamellae grow normal
to this envelope, which for a small perturbation is equivalent to
\begin{equation}
\partial_t y(x,t)\approx -V\partial_x \zeta(x,t),\label{Cahn}
\end{equation}
where $y(x,t)$ is the lateral displacement
of trijunctions from their steady-state positions. Finally, it
follows from the definition of $y$ that
$\lambda(x,t)\approx \lambda_0(1+\partial_x y)$.
Differentiating both sides of this identity with respect
to time and using Eqs. \ref{JH}, \ref{adia}, and \ref{Cahn},
one obtains the diffusion equation (\ref{diff})
with $D=D_\perp$.

We checked that Eq. \ref{adia} is indeed faithfully obeyed in the
range of wavelengths that we consider here.  Consequently, the
discrepancy between $D$ and $D_\perp$ must originate from a correction
to Cahn's normal growth assumption. It is simple to show that
the modified phenomenological form
\begin{equation}
\partial_t y(x,t)\approx -V\partial_x \zeta(x,t)+
      D_\parallel
\partial_x \lambda(x,t)/\lambda_0 ,\label{new}
\end{equation}
yields the diffusion equation (\ref{diff}) with $D$ given by Eq.
\ref{Dnew}. Equation~\ref{new} implies that trijunctions also move
locally ``parallel'' to the envelope of the eutectic interface in
response to a gradient of spacing.  To see physically why this lateral
motion overstabilizes the pattern, consider a local depression in an
array of initial spacing $\lambda_0$.  Cahn's normal growth assumption
implies that such a depression will produce a local decrease of
spacing, and hence a local increase in undercooling that will amplify
this depression if $\lambda_0<\lambda_m$ (because $d\Delta
T/d\lambda<0$ in this case). This well-known argument yields
$\lambda_c=\lambda_m$.  In contrast, the second term on the
right-hand-side of Eq. \ref{new} implies that the lateral motion of
trijunctions opposes the local decrease in $\lambda$, and hence helps
flatten the interface.

A prediction for $\Lambda_c=\lambda_c/\lambda_m$ can be obtained 
by setting $D=D_\perp+D_\parallel=0$, which, using Eqs. \ref{sel1},
\ref{Dperp}, and \ref{Dpar}, yields the cubic equation
\begin{equation}
1-\frac{1}{\Lambda_c^2}+\frac{AG}{K_1V} \,\Lambda_c=0 .\label{pred}
\end{equation}
With $K_{1}= 1.9 \times 10^{-3} Ks\mu m^{-2}$ and $A=0.15$,
we obtain $\Lambda_c = 0.70$
for the experiment of Fig.~\ref{under}; the lowest observed
spacings are just above the predicted stability threshold, 
as it should be. For the phase-field simulations ($K_1=0.1120$),
we find $\Lambda_c=0.942$ for $l_T/l_D=20$ and $\Lambda_c=0.715$
for $l_T/l_D=2$.

Two previous stability analyses have predicted that $\lambda_c$ should
be smaller than $\lambda_m$. The one by Caroli {\it et al.}
\cite{Caroli90}, however, is restricted to a large $G$ limit that
cannot be compared with our results. The other by Chen and Davis
\cite{Chen01} does not have this restriction, but predicts a departure
of $\lambda_c$ from $\lambda_m$ that is about one order of magnitude
smaller than found here and predicted by Eq.~\ref{pred}.
We believe that the lateral motion of the trijunctions is
due to a coupling between the diffusion field and the
non-planar front geometry on the scale of the individual
lamellae; such effects would appear only at higher orders
in the analyses cited above. Therefore, an analytical understanding
of eutectic stability at small spacings from sharp-interface
models largely remains to be developed.

The present results shed light on a number of previous observations.
In metallic eutectics, $K_{1}$ is generally close to $10^{-2} Ks\mu
m^{-2}$ \cite{GuHunt85,TriMas91}, so that the value of $V$ below which
the departure of $\Lambda_c$ from unity becomes significant is of
about $1 ~\mu ms^{-1}$ for $G$ in the $100 Kcm^{-1}$ range.  This may
explain the deviation from the law $\bar{\lambda} \propto V^{-0.5}$,
where $\bar{\lambda}$ is some empirically defined average eutectic
spacing, that has been observed at $V$ lower than about $1~ \mu
ms^{-1}$ in a number of metallic eutectics \cite{RacLesou74}.
Similarly, the overstability due to the
lateral motion of the trijunctions may explain why coupled growth in a
peritectic system has recently been found to be stable \cite{Kurz} in
a situation, analogous to that of eutectics at $\lambda <
\lambda_{m}$, where the interface should be unstable according to the
JH-Cahn stability arguments \cite{Boett74}.
Finally, our results also improve our understanding
of the morphological instability that leads to the formation of
eutectic colonies in the presence of a dilute ternary impurity
\cite{AkaFai00,Plapp01}.

This work was supported by Centre National d'Etudes Spatiales,
France, and by the U.S.~DOE under grant No. DE-FG02-92ER45471.


\begin{thebibliography}{99}

\bibitem{KurFis} W. Kurz and D. J. Fisher, {\sl Fundamentals of
Solidification}, Trans Tech, Aedermannsdorf, Switzerland (1992).

\bibitem{Jackson66} K.~A.~Jackson and J.~D.~Hunt,
  Trans. Metall. Soc. AIME {\bf 236}, 1129 (1966).

\bibitem{Karma96} A.~Karma and A.~Sarkissian,
  Met. Trans. A {\bf 27}, 635 (1996);
  A.~Sarkissian, PhD thesis, Northeastern University, Boston (1996).

\bibitem{Ginibre97} M.~Ginibre, S.~Akamatsu, and G.~Faivre,
  Phys. Rev. E {\bf 56}, 780 (1997).

\bibitem{Langer80} J. S. Langer, Phys. Rev. Lett. {\bf 44}, 1023 (1980).

\bibitem{AkaFai00} S. Akamatsu and G. Faivre, 
  Phys.  Rev.  E {\bf 61}, 3757 (2000).

\bibitem{kassmis} K. Kassner and C. Misbah,
  Phys. Rev. A {\bf 44}, 6513 (1991).

\bibitem{MerFai93} J. Mergy, G. Faivre, C. Guthmann and R. Mellet,
  J.~Cryst.  Growth {\bf 134}, 353 (1993).

\bibitem{Plapp01} M. Plapp and A. Karma, cond-mat/0112194 (2001).

\bibitem{Caroli90} K.~Brattkus, B.~Caroli, C.~Caroli, and B.~Roulet,
  J. Phys. (France) {\bf 51}, 1847 (1990);
  B.~Caroli, C.~Caroli, B.~Roulet, {\sl ibid.} {\bf 51}, 1865 (1990).

\bibitem{Chen01} Y.-J.~Chen and S.~H.~Davis,
  Acta Mater. {\bf 49}, 1363 (2001).
 
\bibitem{GuHunt85} M. G{\"u}nd{\"u}z and J.D. Hunt,
  Acta Metall. {\bf 33}, 1651 (1985).

\bibitem{TriMas91} R. Trivedi, J.~T. Mason, J.~D. Verhoeven, and W. Kurz,
  Metall. Trans. {\bf 22A} 252 (1991).

\bibitem{RacLesou74} R. Racek, G. Lesoult and M. Turpin,
  J.~Cryst. Growth {\bf 22} 210 (1974).

\bibitem{Kurz} M.~Vandyoussefi, H.~W.~Kerr, and W.~Kurz,
  Acta Mater. {\bf 45}, 4093 (1997); S. Dobler {\sl et al.} (unpublished).

\bibitem{Boett74} W.~J.~Boettinger, Metall. Trans. {\bf 5}, 2023 (1974).

\end{thebibliography}
\end{document}